# Classification of synchronization in nonlinear systems using ICO learning

Jyoti Prasad Deka (0000-0002-4110-9294) [1,*]

[1] Department of Physics, Girijananda Chowdhury University, Guwahati-781017, Assam

[*] Corresponding author email ID: jyotiprasad_physics@gcuniversity.ac.in

**Abstract**: In this work, we investigate the implications of the differential Hebbian learning rule known as *Input-Correlations (ICO) learning* [1] in the classification of synchronization in coupled nonlinear oscillator systems. We are investigating the parity-time symmetric coupled Duffing oscillator system with nonlinear dissipation/amplification. In our investigation of the temporal dynamics of this system, it is observed that the system exhibits chaotic as well as quasiperiodic dynamics. On further investigation, it is found that the chaotic dynamics is distorted anti-phase synchronized, whereas the quasiperiodic dynamics is desynchronized. So, on the application of the *ICO learning* in these two parametric regimes, we observe that the weight associated with the stimulus remains constant when the oscillators are anti-phase synchronized, in spite of there being distortion in the synchronization. But when the oscillators exhibit quasiperiodic dynamics, there is erratic evolution of the weight with time. So, from this, it could be ascertained that the *ICO learning* could be made use of in the classification of synchronization dynamics in nonlinear systems.

## 1. INTRODUCTION

In neuroscience, the famous quote '*Neurons that fire together wire together*' explains the synaptic efficacy when the post-synaptic neurons are continuously stimulated by the pre-synaptic neurons [2]. Hebbian learning is one novel attempt that attempts to explain this method in which neurons communicate with each other. It was introduced by Prof. Donald O. Hebb in his book '*The organization of* behavior' [3]. Prof. Hebb emphasized that neurons firing at the same time would not lead to the causality that will strengthen the neural network. Instead, it should be the firing of one neuron thereby leading to the firing of the nearest neuron that will strengthen the neural network.

Spike-timing-dependent plasticity (STDP) which explains the temporal evolution of the strength of communication of neurons with their nearest neighbours is one vital feature of *Hebbian learning* [4-6]. It is an attempt to model two neural processes – Long-Term Potentiation (LTP) or the strengthening of neural connection and Long-Term Depression (LTD) or the weaking of neural connection. In this context, *Input-Correlations* (ICO) *learning* is a method that incorporates the principles of *classical Hebbian Learning* (CHL) together with those of heterosynaptic plasticity [7-8]. It is an unsupervised learning rule put forward by Prof. F. Wörgotter and Prof. B. Porr [1, 17-18], and one major advantage of this method is the replacement of the potentially destabilizing term in the derivative of the output in CHL with the derivative of the input. This enables the use of higher learning rates with no forms of destabilizing in the learning rule.

On the other hand, all forms synchronization – identical, in-phase and anti-phase are exhibited by nonlinear as well as linear systems. It extends from the cosmic to the sub-atomic scale. In the former, the dynamics of the planet in the solar system could be mentioned [9], whereas in the latter, the operating principle of a laser (with billions of atoms pulsating in synchrony) could be talked about as a form of collective synchronization in a physical system [10]. The diffusively coupled *Lorenz* system is one nonlinear system in which identical synchronization is observed when the system is subjected to strong coupling, and it also exhibits synchronization bubbling, which is an intermittent deviation from identical synchronization in certain specific parametric regime of the system [11-12]. In this context, a recently reported work depicts that waveguide coupler could also exhibit identical synchronization, but if the strength of nonlinearity in the system is increased, the system which is initially identically synchronized will start exhibiting loss of any form of synchronization in the system [13]. In non-Hermitian physics, parity-time symmetry (*PT*) is one area in which coupled oscillator systems are observed to exhibit anti-phase synchronized periodic dynamics in the unbroken-*PT* regime of the system, which eventually leads to total loss of synchronization in the broken-*PT* regime thereby implying no correlation between the individual oscillators [14].

*Pearson's correlation coefficient* is one method using which synchronization could be quantified. But the *ICO learning* rule considers the temporal separation between the individual signals and accordingly, the synaptic strength could either increase or decrease and as such, it could be said that this method could be used in the classification of synchronization in nonlinear systems. And this method was also experimentally realized in pulsed optoelectronic setup by Prof. Claudio R. Mirasso, Prof. Ingo Fischer and his group in IFISC-UIB [15]. It was reported that when the stimulus signal precedes the reference, the synaptic strength increases and when the vice-versa occurs, the synaptic strength decreases. But such evolution in the synaptic strength occurs at the exact time-step when the signals overlap. In other words, it could be said that this method is a perfect example of a coincidence detector. In this work, we aim to report about how this rule could be used in the classification of anti-phase synchronization and desynchronization in a *PT*-symmetric nonlinear oscillator model [16]. The model exhibits phase transition from anti-phase synchronized chaotic to desynchronized quasiperiodic dynamics and so, through the application of the *ICO learning* rule in these two parametric regimes, we aim to explain how this rule could be used in the classification of synchronization in nonlinear systems.

In section II, we discuss in detail the mathematical model of the system and also the *ICO learning* rule, and this is followed by the application of this method in the two parametric regimes in section III. And in section IV, we conclude our work with an emphasis on the application of this method in other areas of science and engineering.

## 2. MATHEMATICAL MODEL

The mathematical model of the system is given below.

$$\frac{d^2 x_0}{dt^2} - \gamma x_0^2 \frac{dx_0}{dt} + \alpha x_0 + \beta x_0^3 + \kappa x_1 = 0$$

(1a)

$$\frac{d^2 x_1}{dt^2} + \gamma x_1^2 \frac{dx_1}{dt} + \alpha x_1 + \beta x_1^3 + \kappa x_0 = 0$$

(1b)

Here, $\gamma$ is coefficient of nonlinear dissipation/amplification, $\alpha$ is natural frequency, $\beta$ is strength of Kerr nonlinearity and $\kappa$ is coupling constant. It could be easily verified that this system as whole remains invariant on the joint operation of the parity ($x_0 \leftrightarrow x_1$) and time-reversal operator ($t \rightarrow -t$). Now, before applying the *ICO learning* rule to this mathematical model, we shall first try to understand the *ICO learning* rule and how it could be used in the classification of temporal dynamics in coupled nonlinear systems. So, at first, we shall try to evaluate the temporal maxima of the time-series of the functions below.

$$x_0(t) = \sin(\omega_0 t)$$
$$x_1(t) = \sin(\omega_0 t + \phi)$$

Here, $x_0(t)$ is the reference and $x_1(t)$ is the stimulus signal. $\omega_0 = 1$ and $\phi = \pm \pi/25$ are the frequency and the temporal delay of the signals respectively. After the evaluation of the temporal maxima of these signals, we need to use the function given below to temporally extend the signals so that overlap in the signals could be achieved.

$$f(n) = \frac{c}{\sqrt{(2\pi f)^2 - (\pi f/Q)^2}} e^{-\pi fn/Q} \sin\sqrt{(2\pi f)^2 - (\pi f/Q)^2}\, n$$

(2)

Here, $n$ is discrete time-step, $c$ is amplitude, $f$ is frequency and $Q$ is decay rate. In our simulation, we used the following parameters - $f = 0.01$, $Q = 0.51$ and $c = 1$. Now after obtaining the temporally extended signals, the weight $w_0$ associated with the reference is kept constant at $w_0 = 1$ and the weight $w_1$ (initialized at $w_1 = 1$) associated with the stimulus is calculated using the equation below.

$$\frac{dw_1}{dt} = \eta u_1 \frac{du_0}{dt}$$

(3)

Here, $\eta = 0.01$ is the learning rate and $u_i$ are the temporally extended signals. And from the evaluation of the weights, the output of the *ICO learning* rule is given by

$$v = w_0 u_0 + \sum_{i=1}^{n} w_i u_i$$

(4)

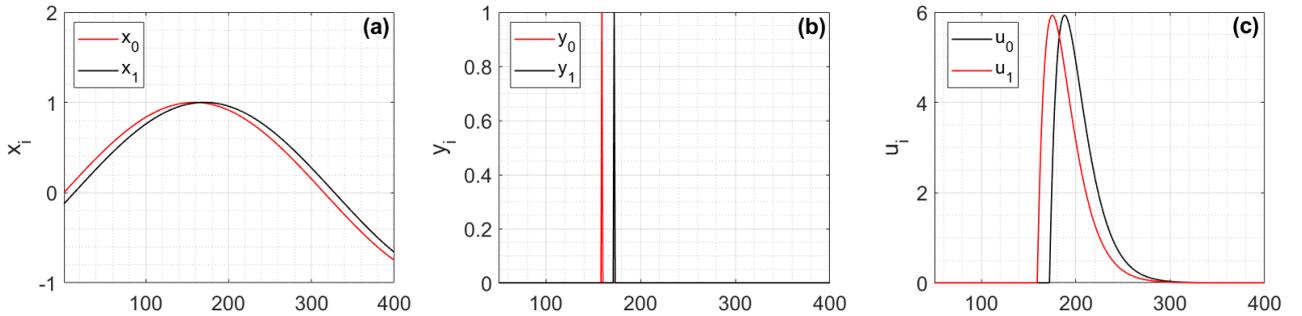

**Fig. 1**. Temporal evolution of the (a) continuous time sinusoidal signals $x_i(t)$, its temporal maxima $y_i(t)$ and (c) the temporally extended signals $u_i(t)$ for $\phi = \pi/25$. Parameters: $f = 0.01$, $Q = 0.51$ and $c = 1$.

In Fig. 1 above, we present the situation when we use the signals given above in Eq. 1 for $\phi = \pi/25$. It could be seen that the stimulus (**solid red**) precedes the reference (**solid black**) for the chosen parameters and the temporally extended signals is spiking initially followed by an exponential decay with time. It should be noted that the temporally extended signals exist when a temporal maximum in the continuous-time sinusoidal signals is evaluated. In Fig. 2, we present the temporal evolution of the weights $w_i$ and it could be seen that the weight $w_1$ associated with the stimulus increases with time and the spiking increment observed in the temporal evolution is observed at the exact time-step when the temporally extended signals overlap. From this, it could be ascertained that the *ICO learning* rule also acts as a coincidence detector in continuous-time signals and the maxima of the spiking output has been observed to be increasing with time in accordance with the temporal increment of $w_1$. Now, in Fig. 3, we present the temporal evolution of the weights when the reference precedes the stimulus for $\phi = -\pi/25$. Here, the exact opposite is observed. $w_1$ decreases with time and so does the output $v$. In fact, no spiking behavior in the temporal evolution of $w_1$ is observed. In other words, these two specific circumstances presented in Fig. 2 and Fig. 3 delineate the circumstances in which the synaptic strength in neural networks strengthens and weakens with time, and this is what is termed as plasticity in a neural network.

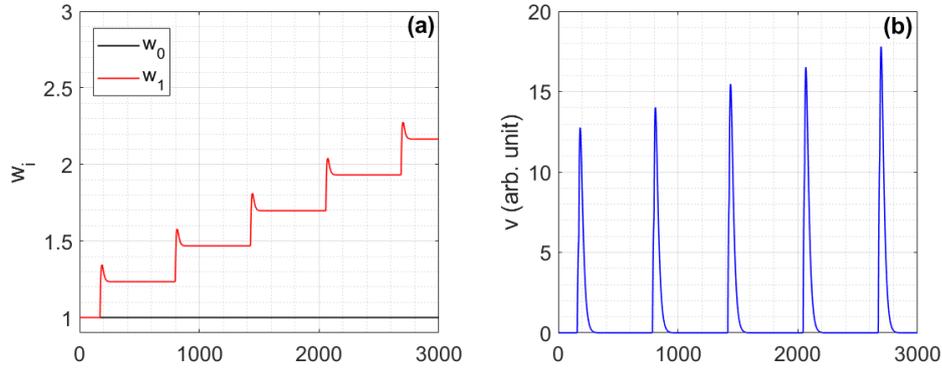

**Fig. 2**. (a) Temporal evolution of the weights $w_i$ and (b) Output of the *ICO learning* rule $v$ when the stimulus precedes the reference for $\phi = \pi/25$ and $\eta = 0.01$.

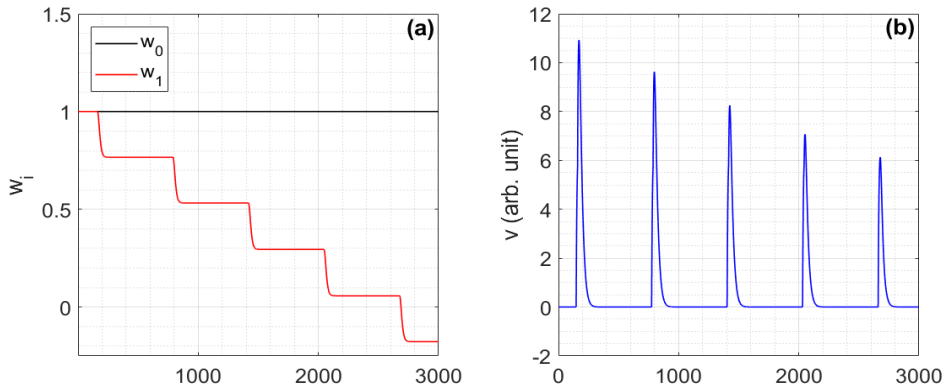

**Fig. 3**. (a) Temporal evolution of the weights $w_i$ and (b) Output of the *ICO learning* rule when the reference precedes the stimulus for $\phi = -\pi/25$ and $\eta = 0.01$.

## 3. RESULTS AND DISCUSSION

We shall now discuss how the *ICO learning* rule could be used in the classification of synchronization in nonlinear systems. In Fig. 4, we present two specific cases of the mathematical model presented in Eq.1. In Fig. 4a and 4c, it could be seen that the gain oscillator exhibits chaotic dynamics for $\alpha = 0.25$ and for $\alpha = 0.55$, the same oscillator exhibits quasiperiodic dynamics. The phase plane figures shown in Fig. 4 are a signature of the chaotic and quasiperiodic temporal evolution of the two systems and at the same time, the plot of $x_1$ v/s. $x_0$ shown in Fig. 4b depicts that the chaotic dynamics is distorted anti-phase synchronized and the same shown in Fig. 4d for $\alpha = 0.55$ depicts the desynchronized quasiperiodic dynamics in the two systems. In short, it could be said that there is a phase transition in the synchronization dynamics of the oscillators as $\alpha$ is varied [15].

On the application of the *ICO learning* rule in the time-series of the nonlinear system in these two parametric regimes, we observe from Fig. 5 that when the oscillators are anti-phase synchronized, $w_1$ remains constant and when they are desynchronized, it varies in an erratic manner. In the previous section, we considered two sinusoidally varying signals with temporal delay between them, and we observed that the $w_1$ increases or decreases with time subjected to the temporal delay between the two signals. But here, we observed that the weight $w_1$ remains static when the two oscillators are anti-phase synchronized and when there is absence of any form of synchronization, $w_1$ behaves in an erratic manner with no predictability in its evolution. This means when the two oscillators are anti-phase synchronized, there is no overlap between the two temporal extended signals of the system and when there is no synchronization between them, the signals overlap intermittently with no uniformity in its evolution.

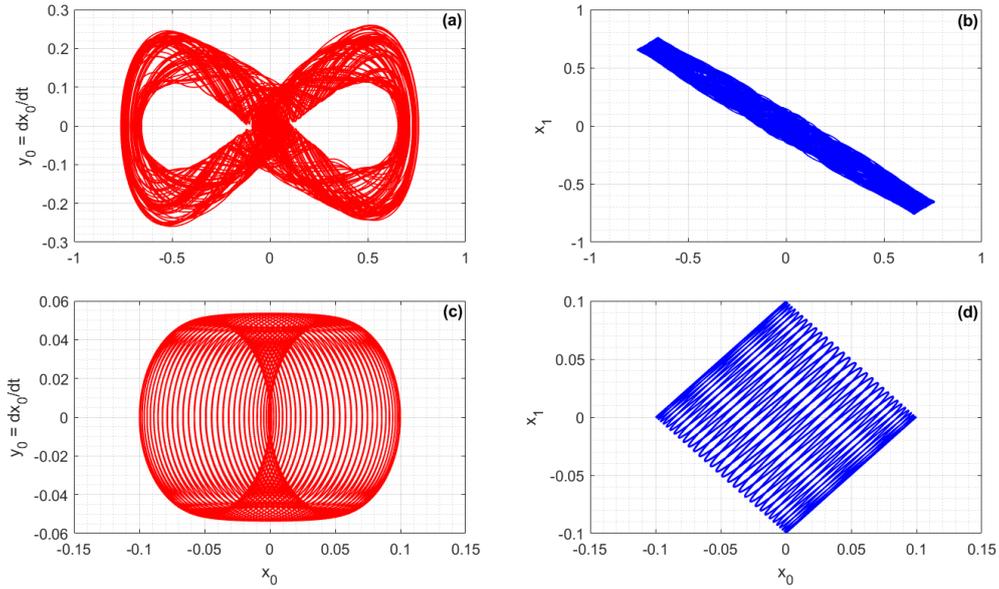

**Fig. 4**. (a), (c) Phase plane of the gain oscillator and (b), (d) $x_1$ v/s. $x_0$ for $\alpha = 0.25$ and $\alpha = 0.55$ respectively.

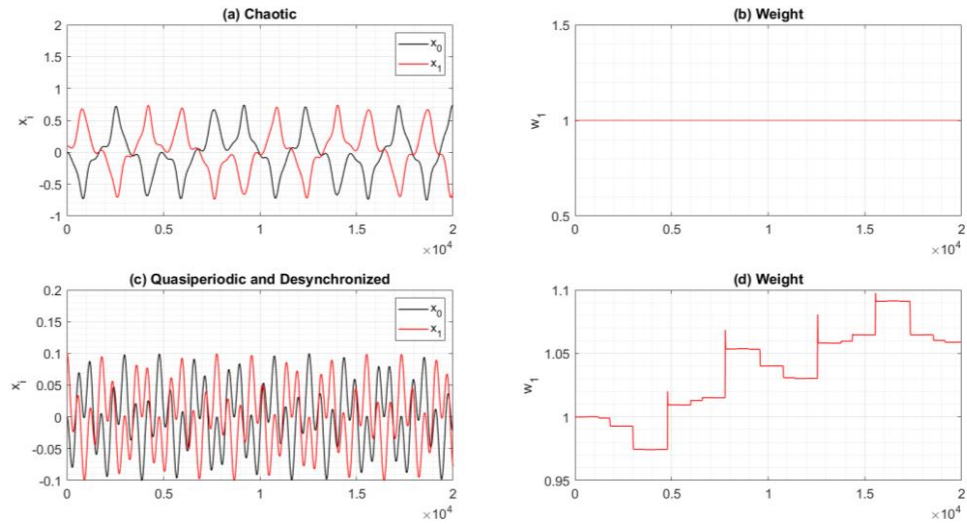

**Fig. 5**. (a), (c) Temporal evolution of the oscillators and (b), (d) the Dendritic weight $w_1$ associated with the stimulus for $\alpha = 0.25$ and $\alpha = 0.55$ respectively. Other parameters - $\eta = 0.01, Q = 0.51, c = 1$ and $f = 0.025$.

Now, it may be argued that if we interchange the two oscillators, will it be possible to observed completely static $w_1$ for anti-phase synchronized dynamics in a coupled oscillator system? To answer this, we would like to point out that the static evolution of $w_1$ in Fig. 5b is an indication of the fact that there is no overlap in the temporally extended signals $u_i$ of the system. So, irrespective of which oscillator we chose as the reference or the stimulus, the static temporal evolution of $w_1$ will be observed in the application of the *ICO learning* rule in time-series classification of anti-phase synchronized systems. Moreover, the discussion in section II further threw light on the fact that using this method, we may observe increasing or decreasing temporal evolution of the weight and in that scenario, the two signals are in-phase synchronized. And in Fig. 5d, we showed how the weight varies erratically when there is no synchronization of any form in the coupled oscillator system. Hence, we can say that *ICO learning* rule could be used to classify the synchronization dynamics in nonlinear systems.

## 4. CONCLUSION

In conclusion, it could be said that the *ICO learning* rule could be used in the classification of time-series signals of coupled oscillator systems, especially in their synchronization dynamics. In this work, we found that for anti-phase synchronized temporal dynamics, the weight associated with the stimulus exhibits static dynamics and for desynchronized quasiperiodic dynamics, it behaves in an erratic fashion with no predictability. From our analysis, it could be said that this method could also be used in the characterization of time-series in nonlinear systems. Furthermore, our model consists of two oscillators – one of which is the reference and the other the stimulus. It could also be used in oscillator ensembles in which one reference oscillator drives an ensemble of coupled oscillator systems and temporal dynamics such as the emergence of chimera states in the ensemble system could be quantified.

## References

[1] B. Porr, F. Wörgötter, Strongly Improved Stability and Faster Convergence of Temporal Sequence Learning by Using Input Correlations Only, Neural Computation 18 (6), 1380–1412 (2006).
[2] S. Löwel, W. Singer, Selection of intrinsic horizontal connections in the visual cortex by correlated neuronal activity, Science 255, 209–12 (1992).
[3] D. O. Hebb, The Organization of Behavior: A Neuropsychological Theory (1st ed.). Psychology Press. https://doi.org/10.4324/9781410612403 (2002).
[4] M. M. Taylor, The Problem of Stimulus Structure in the Behavioural Theory of Perception, South African Journal of Psychology 3, 23–45 (1973).
[5] W. B. Levy, O. Steward, Temporal contiguity requirements for long-term associative potentiation/depression in the hippocampus, Neuroscience 8, 4, 791–797 (1983).
[6] R. Fitzsimonds, M. Poo, Retrograde Signaling in the Development and Modification of Synapses", Physiological Reviews, 78, 143 (1998).
[7] N. Caporale, Y. Dan, Spike timing-dependent plasticity: a Hebbian learning rule, Annual Review of Neuroscience 31, 25–46 (2008).
[8] O. Paulsen, T. Sejnowski, Natural patterns of activity and long-term synaptic plasticity, Current Opinion in Neurobiology 10 (2), 172–180 (2000).
[9] N. Scafetta, The complex planetary synchronization structure of the solar system, Pattern Recogn. Phys., 2, 1-19 (2014).
[10] R. Roy and K. S. Thornburg Jr., Experimental synchronization of chaotic lasers, Phys. Rev. Lett. 72, 2009 (1994).
[11] E. Park, M. A. Zaks, and J. Kurths, Phase synchronization in the forced Lorenz system, Phys. Rev. E 60, 6627 (1999).
[12] L. M. Pecora, T. L. Carroll, Synchronization of chaotic systems, Chaos 25, 097611 (2015).
[13] J. P. Deka, Onset of identical synchronization in the spatial evolution of optical power in a waveguide coupler, Indian J Phys https://doi.org/10.1007/s12648-024-03348-4 (2024).
[14] C. M. Bender, B. K. Berntson, D. Parker, E. Samuel, Observation of phase transition in a simple mechanical system, Am. J. Phys. 81 (3), 173–179 (2013).
[15] S. Ortín et al., Implementation of input correlation learning with an optoelectronic dendritic unit, Front. in Phys. 11, 1112295 (2023).
[16] J. P. Deka, Anti-Phase Synchronization of Chaos in PT-Symmetric Nonlinear Oscillators, arXiv:2310.12154v1 (2023).
[17] B. Porr, F. Wörgötter, Isotropic sequence order learning, Neural Comput. 15, 831 (2003).
[18] K. Möller, D. Kappel, M. Tamosiunaite, C. Tetzlaff, B. Porr, F. Wörgötter, Differential Hebbian learning with time-continuous signals for active noise reduction, PLoS ONE 17, 5 (2022).